\documentclass[12pt]{article}

\usepackage{amsmath,amsfonts,amssymb}

\def\bA{\mathbf{A}}

\def\be{\begin{equation}}

\def\ee{\end{equation}}

\def\bea{\begin{eqnarray}}

\def\eea{\end{eqnarray}}

\def\mV{\mathcal{V}}

\def \bA{\mathbf{A}}

\def\bB{\mathbf{B}}

\newcommand{\mL}{\mathcal{L}}

\begin{document}

	\begin{titlepage}

		\vskip 0.4 cm
		
		\begin{center}
			{\Large{ \bf Proposal for D0-brane anti-D0-brane Action
			}}
			
			\vspace{1em}  
			
			\vspace{1em} J. Kluso\v{n} 			
			\footnote{Email addresses:
				klu@physics.muni.cz (J.
				Kluso\v{n}) }\\
			\vspace{1em}
			\textit{Department of Theoretical Physics and
				Astrophysics, Faculty of Science,\\
				Masaryk University, Kotl\'a\v{r}sk\'a 2, 611 37, Brno, Czech Republic}
			
			\vskip 0.8cm
			
			%
			%
			%
			%
			%
			%
			
			\vskip 0.8cm
			
		\end{center}

		\begin{abstract}
We propose an action for the D0-brane anti-D0-brane system that has its exact solution 
corresponding to the marginal tachyon profile defined for arbitrary constant separation 
of these D-branes. We also find its covariant formulation and generalization to higher
dimensional Dp-branes.

		\end{abstract}
		
		\bigskip
		
	\end{titlepage}
	
	\newpage

\section{Introduction and Summary}\label{first}
The dynamics of Dp-branes with opposite charges is one of the most interesting 
processes in string theory.  In case, when their separation is larger than string scale, 
their instability can be interpreted as a consequence of the gravity attraction together
with the attractive force mediated by Ramond-Ramond $p+1$-forms that arises between two
branes with opposite charges. In case when their separation is smaller than string length
scale the instability of this system can be interpreted as emergence of the tachyon field 
\cite{Sen:1998ii,Sen:1999mg}. However these modes become massive when the separation of 
Dp-branes is larger than string scale. As was nicely argued in \cite{Garousi:2004rd}
the effective action should depend on the brane separation. When this separation is larger 
than the string scale the effective action should be in terms of massless 
closed string states that propagate between these branes. On the other hand when their separation is smaller than the string scale the more appropriate description of this system should be in terms of the tachyon effective action that captures dynamics when the separation is below string scale. The first proposal for the action that describes dynamics of Dp-brane anti-Dp-brane system was given in 
\cite{Sen:2003tm} that was mainly motivated by the effective actions for non-BPS Dp-branes in string theory \cite{Sen:1999md,Garousi:2000tr,Bergshoeff:2000dq,Kluson:2000iy}. Coincident Dp-brane anti-Dp-brane system was also analysed with the help of boundary string field theory  \cite{Kraus:2000nj,Takayanagi:2000rz}. On the other hand when their separation is larger than string scale the Dp-brane and anti Dp-brane are attracted by attractive force due to the gravity and RR fields.

Very interesting proposal for Dp-brane anti-Dp-brane effective action was given in 
\cite{Garousi:2004rd} where the action for this system was derived by $(-1)^{F_L}$ projection of non-abelian action for two non-BPS Dp-branes \cite{Garousi:2000tr}
that was generalization of non-abelian DBI action \cite{Myers:1999ps}. The resulting
action contains additional terms with respect to the action  introduced in 
\cite{Sen:2003tm}. On the other hand different approach for the study of Dp-brane anti-Dp-brane system at finite distance was performed in \cite{Israel:2011ut}. It was argued in this paper  that there is a marginal tachyon deformation for all values of relative
distance below critical distance. This tachyon deformation corresponds to half-S-brane
solution 
in terminology of the time dependent tachyon deformation on the world-volume of
unstable Dp-brane 
\cite{Sen:2002nu,Sen:2002in,Lambert:2003zr,Kutasov:2003er}. Then it was shown in 
\cite{Israel:2011ut} that the effective action proposed in 
\cite{Garousi:2004rd} cannot have the marginal tachyon condensation for constant 
relative distance as its exact solution. On the other hand leading order form of new tachyon effective action for Dp-brane anti-Dp-brane system was suggested in \cite{Israel:2011ut} that has tachyon marginal profile as exact solution for constant relative distance. 

This suggested
Lagrangian will be the starting point for our proposed effective action for D0-brane anti-D0-brane system that is valid for any value of the tachyon field.  We determine the form of this Lagrangian using following procedure. As the first step we consider tachyon effective action that
has similar structure as the Lagrangian given in \cite{Kutasov:2003er} and which has half S-brane tachyon condensation as its exact solution. Then we show that the marginal tachyon profile that describes  tachyon condensation on D0-brane anti-D0-brane system at finite subcritical distance is  solution of the equations of motion that arises from this proposed naive form of the tachyon effective action. On the other hand when we derive equations of motion for relative distance
we find that they cannot be solved by constant relative distance. For that reason we modify the proposed naive form of the effective action when we add to it term linear in 
the time derivative of relative distance. Then  the tachyon marginal profile at constant relative distance is still its exact solution. However the presence of this new term will modify equations of motion for relative distance when 
now the requirement that the marginal tachyon profile together with the constant relative distance are solutions of equations of motion determines the form of the new term. In other words we find an exact form of the effective action for D0-brane anti-D0-brane system that correctly describes 
its dynamics when their separation is smaller than the critical distance. This action also reduces into the tachyon effective action for non-BPS Dp-brane when we perform $(-1)^{F_L}$ projection which is nice consistency check of our proposal. 

Having derived the effective action for real tachyon and for the relative distance we then determine its covariant form  introducing coordinates that parametrize transverse positions of D0-branes together with world-line gauge field. As a result we derive full form of D0-brane anti-D0-brane system in the flat target space background. Then it is easy task to generalize this form to the case of Dp-brane anti-Dp-brane system. 

Let us outline our results and suggest possible extension of this work. We propose an action for D0-brane anti-D0-brane tachyon effective action that has tachyon marginal profile at constant relative distance as its exact solution at any value of the tachyon field and for the relative distance below the critical distance. This is very important result that shows that tachyon effective action can describe dynamics of such a complex system as D0-brane anti-D0-brane is which is in agreement with the general definition of the tachyon effective action. Then we generalized this action to the case of Dp-brane anti-Dp-brane system. It is also important to stress that 
the action is not unique. In fact, it would be very important to study this action further and try to see whether it is related to the action proposed in \cite{Garousi:2004rd}. It is clear that they cannot be identified but rather they should be related by some redefinition of the tachyon field in the similar way as the tachyon effective action given in \cite{Kutasov:2003er} is related to the DBI like tachyon effective action suggested in \cite{Sen:1999md,Garousi:2000tr,Bergshoeff:2000dq,Kluson:2000iy}. We hope to return to these problem in future. 

This paper is organized as follows. In the next section (\ref{second}) we review basic facts about effective action for Dp-brane anti-Dp-brane system. Then in section (\ref{third}) we present construction of the new action for this system. Finally in section (\ref{fourth}) we determine covariant form of this action and present its generalization to Dp-brane anti-Dp-brane system.

\section{Review of Standard Formulation of  $Dp-\overline{D}p$-Brane Action}
\label{second}
In this section review basic facts about effective action that describes Dp-brane anti-Dp-brane system. 
 The first proposal for the action that describes
this system was given in \cite{Sen:2003tm} and it has the form
\begin{equation}
S=-\int d^{p+1}\xi V(T,X^{(1)}-X^{(2)})
(\sqrt{-\det \bA_{(1)}}+\sqrt{-\det \bA_{(2)}}) \ ,
\end{equation}
where 
\begin{eqnarray}
& |&\bA_{(i)\alpha\beta}=\eta_{\alpha\beta}+\partial_\alpha X^{(i)I}\partial_\beta X^{(i)J}\delta_{IJ}+
\lambda F_{\alpha\beta}^{(i)}+\lambda\frac{1}{2}((D_\alpha T)^*D_\beta T+
\frac{1}{2}(D_\beta T)^*D_\alpha T) \ , \nonumber \\
& &F_{\alpha\beta}^{(i)}=\partial_\alpha A_\beta^{(i)}-\partial_\beta A_\alpha^{(i)} \ , \quad 
D_\alpha T=(\partial_\alpha -iA_{\alpha}^{(1)}+iA_\alpha^{(2)})T \ , \nonumber \\
\end{eqnarray}
and where the tachyon potential has the form 
\begin{equation}
V(T,X^{(1)}-X^{(2)})=
T_p[1+\frac{1}{2}(\sum_I\frac{(X_{(1)}^I-X_{(2)}^I)^2}{\lambda})T^2+O(T^4)] \ , 
\end{equation}
where $T_p$ is Dp-brane tension defined as
\begin{equation}
T_p=\frac{2\pi}{(2\pi l_s)^{p+1}}=\frac{1}{2\pi^{\frac{2p+1}{2}}\alpha'^{\frac{p+1}{2}}} \ ,     \quad l_s^2=2\pi\alpha'\equiv \lambda \ . 
\end{equation}
Let us explain our notation and convention. $\xi^\alpha,\alpha,\beta=0,1,\dots,p$ 
label world-volume of Dp-branes where we presume static gauge 
\begin{equation}
	\xi^\alpha=X^{(1)\alpha}=X^{(2)\alpha}  \ . 
\end{equation}
Note that Dp-brane anti-Dp-brane pair are extended in flat Minkowski space-time 
with the target space metric $\eta_{\alpha\beta}=\mathrm{diag}(-1,1,\dots,9)$ where the transverse positions of Dp-brane and anti-Dp-brane are labelled with coordinates $X^{(i)I}$, where $i=1$ corresponds to Dp-brane and $i=2$ corresponds to anti-Dp-brane and where $I=p+1,\dots,9$. 
Finally, $A_{\alpha}^{(1)}$ is the gauge field living on the world-volume of Dp-brane and 
$A_\alpha^{(2)}$ living on anti-Dp-brane. 

The nice generalization of this action was presented in \cite{Garousi:2004rd}(see also 
\cite{Garousi:2007fn}) where this action was derived from an effective action for two non-BPS Dp-branes by projecting it with $(-1)^{F_L}$ where $F_L$ is space-time left handed fermion number. In the flat background the resulting action has the form 
\begin{equation}
S=-\int d^{p+1}\xi \mV(T,r)\left(\sqrt{-\det \bA_{(1)}}+
\sqrt{-\det \bA_{(2)}}\right) \ , 
\end{equation}
where 
\begin{equation}
\mV(T,r)=V(|T|)\sqrt{1+\frac{T^2}{\lambda}r^Ir_I} \ , \quad r^I=X^{(1)(I)}-X^{(2)I} \ ,  
\end{equation}
and where $\bA_{(i)}$ has more general form
\begin{eqnarray}
& &\bA^{(i)}_{\alpha\beta}=\eta_{\alpha\beta}+
\delta_{IJ}\partial_\alpha X^{(i)I}\partial_\beta X^{(i)J}
-\frac{TT^*}{\lambda(1+\frac{TT^*}{\lambda}r^Ir_I)}
\partial_\alpha X^{(i)I}\delta_{IJ}r^Jr^K\delta_{KL}\partial_\beta X^{(i)L}
+\lambda F_{\alpha\beta}^{(i)}+
\nonumber \\
&&+\frac{1}{1+\frac{TT^*}{\lambda}r^Ir_I}
\frac{\lambda}{2}(D_\alpha T(D_\beta T)^*+D_\beta T(D_\alpha T)^*)+\nonumber \\
&&+\frac{i}{2}\partial_\alpha X^{(i)I}\delta_{IJ}r^J
(T(D_\beta T)^*-T^* (D_\beta T))
-\frac{i}{2}(T(D_\alpha T)^*-T^* D_\alpha T)r^I\delta_{IJ}\partial_\beta X^{(n)J} \ . 
\nonumber \\
\end{eqnarray}
Following \cite{Garousi:2004rd} we can gain more physical insight into this action  when consider pure time dependent situation and presume that Dp-branes are separated in the one direction only. We further write the tachyon field as $T=\tau e^{i\rho}$ and also 
write $X^{(1)}=R+\frac{r}{2} \ , X^{(2)}=R-\frac{r}{2}$. Then it is 
easy to see that  the effective action does not depend on $R$ so that  without lost of generality we  take $R=\mathrm{const}$. Then we have
\begin{eqnarray}
\bA^{(1)}_{00}=\bA_{00}^{(2)}=-1+\frac{1}{4}\dot{r}^2-
\frac{\tau^2}{4\lambda(1+\frac{\tau^2}{\lambda}r^2)}\dot{r}^2r^2+
\frac{\lambda}{(1+\frac{\tau^2}{\lambda}r^2)}(\dot{\tau}^2+\tau\dot{\rho}^2) \ , \nonumber \\
\end{eqnarray}
so that the Lagrangian density takes the form
\begin{equation}
\mL=-2 V(\tau^2)\sqrt{1+\frac{1}{\lambda}\tau^2 r^2-\frac{1}{4}\dot{r}^2-
\lambda\tau^2-\lambda \tau^2\dot{\rho}^2} \ . 
\end{equation}
The properties of this Lagrangian were very extensively studied in 
\cite{Garousi:2004rd}. 
On the other hand it was argued in 
\cite{Israel:2011ut} that this Lagrangian does not capture an important exact solution of the tachyon condensation corresponding to the profile
\begin{equation}\label{taumarg}
\tau_{marg}=\tau_0 \exp \frac{\omega}{\sqrt{\lambda}}\xi^0 \ ,
\end{equation}
where $\omega$ is given by condition
\begin{equation}
\omega^2=\frac{1}{2}-\frac{r^2}{\lambda} \ .
\end{equation}
In more details, let us consider the leading order form of Dp-brane anti-Dp-brane Lagrangian  that was proposed in  
\cite{Israel:2011ut}
\begin{equation}\label{mLIsrael}
	\mL=-2+\sqrt{1-2\lambda^{-1}r^2}\left(\frac{\tau^2}{2}+\lambda\frac{\dot{\tau}^2}{1-2\lambda^{-1}r^2}\right) \ . 
\end{equation}
The equations of motion for $r$ and $\tau$ that follow from this Lagrangian, have the form
\begin{eqnarray}
& &-\frac{2r\lambda^{-1}}{\sqrt{1-2\lambda^{-1}r^2}}
\left(\frac{\tau^2}{2}+\lambda\frac{\dot{\tau}^2}{1-2\lambda^{-1}r^2}\right)
+\sqrt{1-2\lambda^{-1}r^2}\frac{4\lambda^{-1}r}{(1-2\lambda^{-1}r^2)^2}\lambda
\dot{\tau}^2=0 \ , 
\nonumber \\
& &\tau-2\lambda\frac{d}{d\xi^0}\left(\frac{\dot{\tau}}{1-2\lambda^{-1}r^2}\right)=0 \ . 
\nonumber \\
\end{eqnarray}
Then using the fact that for $\tau_{marg}$ we have $\dot{\tau}_{marg}=\frac{\omega}{\sqrt{\lambda}}\tau$ we find that two equations given above are obeyed for $\tau_{marg}$ and for $r=\mathrm{const}$.

 The goal of this paper is to find possible generalization of the Lagrangian density (\ref{mLIsrael}) for any value of the tachyon field that has marginal solution as its exact solution in the similar way how the tachyon effective action for unstable Dp-brane has half S-brane solution as its exact solution. This was nicely shown  in \cite{Kutasov:2003er} \footnote{For related work, see 
for example \cite{Kluson:2004ns,Kluson:2004qy}.} where it was also shown that full
S-brane tachyon profile is exact solution of corresponding equations of motion.  
 The construction of such an action for D0 anti-D0-brane system will be presented in the next section.
\section{Proposal for D0-brane anti-D0-brane Action}\label{third}
In this section we propose form of D0-brane anti-D0-brane action.

We start our construction with the observation that the Lagrangian density (\ref{mLIsrael}) can be written in the leading order as 
\begin{eqnarray}\label{mLnaive}
&&	\mL=-2T_0\left(1-\sqrt{1-2\lambda^{-1}r^2}\frac{\tau^2}{4}-\lambda\frac{\dot{\tau}^2}{2\sqrt{1-2\lambda^{-1}r^2}}\right)=\nonumber \\
&&	=-2T_0(1-\sqrt{1-2\lambda^{-1}r^2}\frac{\tau^2}{2})(1+\frac{\tau^2}{4}\sqrt{1-2\lambda^{-1}r^2}-\lambda\frac{\dot{\tau}^2}{2
		\sqrt{1-2\lambda^{-1}r^2}})\approx\nonumber \\
&&\approx-2T_0\frac{1}{1+\sqrt{1-2\lambda^{-1}r^2}\frac{\tau^2}{2}}
\sqrt{1+\frac{\tau^2}{2}\sqrt{1-2\lambda^{-1}r^2}-\lambda\frac{\dot{\tau}^2}{
	\sqrt{1-2r^2}}}	\ , 
	\nonumber \\
\end{eqnarray}
where overall pre factor $2$ comes from the fact that this action describes two D0-branes  and in the last step we extrapolated the expression  in the second bracket to the
square root structure. Then the last form of the action for $r=0$ reduces to the tachyon effective action for D0-brane in Type IIB theory \cite{Kutasov:2003er}
\begin{equation}
\mL_{non}=-\frac{T_{0}^{non}}{1+\frac{\tau^2}{2}}
	\sqrt{1+\frac{\tau^2}{2}-\lambda\dot{\tau}^2}	\ . 
\end{equation}
where $T_{0}^{non}=\sqrt{2}T_0$. This is the first important sign how the effective action for 
D0-brane anti-D0-brane system should have the form since as was argued in 
  \cite{Sen:2003tm} this action should reduce for $r=0$ to the action for unstable D0-brane. In fact, the original unstable D0-brane was defined as moding out D0-brane anti-D0-brane by 
  $(-1)^{F_L}$ operation \cite{Sen:1999mg}.

To proceed further we introduce  $\bB$ defined as
\begin{equation}
	\bB=1+\frac{\tau^2}{2}\sqrt{1-2\lambda^{-1}r^2}-\lambda\frac{\dot{\tau}^2}{
		\sqrt{1-2\lambda^{-1}r^2}}
\end{equation}
that has the property that for $\tau=\exp \omega \frac{\xi^0}{\sqrt{\lambda}}$ it is equal to
\begin{equation}
	\bB=1+\frac{\tau^2}{2}\sqrt{1-2\lambda^{-1}r^2}-\frac{\omega^2\tau^2}{\sqrt{1-2\lambda^{-1}r^2}}
\end{equation}
and this is equal to $1$ when $\omega$ is equal to 
\begin{equation}
	 \omega^2=\frac{1}{2}-\frac{r^2}{\lambda} \ .
\end{equation}
This is condition of marginality of the tachyon profile for constant $r$
\cite{Israel:2011ut}.

Then we should ask the question whether the form of the Lagrangian density given 
on the last line on (\ref{mLnaive}) could be the right one.  Explicitly, let us denote this Lagrangian density as
\begin{equation}\label{mLnaive2}
		\mL_{naiv}=-\frac{2T_0}{1+\frac{\tau^2}{2}\sqrt{1-2\lambda^{-1}r^2}}
		\sqrt{\bB} \ . 
\end{equation}
Now we should explicitly show
that the equations of motion for $r$ and $\tau$ that follow from (\ref{mLnaive2}) have tachyon 
marginal profile $\tau_{marg}$ and $r=\mathrm{const}$ as exact solutions.  First of all the equation 
of motion for $\tau$ that follow from 
(\ref{mLnaive2}) has the form
\begin{eqnarray}\label{eqtaunaive}
&&-\frac{\tau\sqrt{1-2\lambda^{-1}r^2}}{(1+\frac{\tau^2}{2}\sqrt{1-2\lambda^{-1}r^2})^2}
\sqrt{\bB}+\frac{\sqrt{1-2\lambda^{-1}r^2}}{1+\frac{\tau^2}{2}\sqrt{1-2\lambda^{-1}r^2}}\frac{\tau}{2\sqrt{\bB}}
\nonumber \\
&&+\frac{d}{d\xi^0}\left(\frac{1}{1+\frac{\tau^2}{2}\sqrt{1-2\lambda^{-1}r^2}}
\frac{\lambda\dot{\tau}}{\sqrt{\bB}\sqrt{1-2\lambda^{-1}r^2}}\right)=0 \ . 
\nonumber \\
\end{eqnarray}
Then inserting $\tau_{marg}$ given in (\ref{taumarg}) into (\ref{eqtaunaive}) we get
\begin{eqnarray}
-\frac{\tau}{(1+\frac{\tau^2}{2}\sqrt{1-2\lambda^{-1}r^2})^2}
(\sqrt{1-2\lambda^{-1}r^2}+\omega^2\tau^2)+
\frac{\tau}{1+\frac{\tau^2}{2}\sqrt{1-2\lambda^{-1}r^2}}\sqrt{1-2\lambda^{-1}r^2}=0 \ \nonumber \\
\end{eqnarray}
that shows that $\tau_{marg}$ solves the equation of motion which suggests that 
(\ref{mLnaive2}) could be step in the right direction. To proceed further let us  study equation of motion for $r$ that follows from (\ref{mLnaive2})
\begin{eqnarray}\label{eqrnaive}
&&	-\frac{1}{(1+\frac{\tau^2}{2}\sqrt{1-2\lambda^{-1}r^2})^2}\frac{\tau^2 r}{\sqrt{1-2\lambda^{-1}r^2}}\sqrt{\bB}
\nonumber \\
&&	+\frac{1}{1+\frac{\tau^2}{2}\sqrt{1-2\lambda^{-1}r^2}}\frac{1}{\sqrt{\bB}}
	\frac{r }{\sqrt{1-2\lambda^{-1}r^2}}
\left(\frac{\tau^2}{2}+\frac{\lambda\dot{\tau}^2}{(1-2\lambda^{-1}r^2)}\right)=0 \ , 
	\nonumber \\
\end{eqnarray}
where we used 
\begin{equation}
	\frac{\delta \sqrt{\bB}}{\delta r}=
	-\frac{1}{\sqrt{\bB}}
	\frac{\lambda^{-1}r }{\sqrt{1-2\lambda^{-1}r^2}}
	\left(\frac{\tau^2}{2}+\frac{\lambda\dot{\tau}^2}{1-2\lambda^{-1}r^2}\right)	\ . 
\end{equation}
Inserting (\ref{taumarg}) into (\ref{eqrnaive})  we find that this equation is equal to
\begin{eqnarray}
&&\frac{r\tau^4}{2(1+\frac{\tau^2}{2}\sqrt{1-2\lambda^{-1}r^2})^2}
\nonumber \\
\end{eqnarray}
that shows that the solution $r=\mathrm{const}$ and $\tau_{marg}$ are not solutions of the equation of motion (\ref{eqrnaive}). In other words we should modify (\ref{mLnaive2}) in such way that $\tau_{marg}$ and $r=\mathrm{const}$ will be solutions of the equations of motion for $\tau$ and $r$. For that reason we add new term to  Lagrangian density  and we demand that the presence of this new term  does not change equation of motion for $\tau$ in case when $r=\mathrm{const}$. Explicitly, we add term linear in $\dot{r}$ to  $\bB$ so that
\begin{equation}\label{Bmod}
\bB=1+\frac{\tau^2}{2}\sqrt{1-2\lambda^{-1}r^2}-2f(\tau,r)\dot{r}r\dot{\tau}
-\lambda\frac{1}{\sqrt{1-2\lambda^{-1}r^2}}\dot{\tau}^2 \ 
\end{equation}
that is again equal to $1$ for $\tau=\tau_{marg}$ and for $r=\mathrm{const}$. Note that $f(\tau,r)$ is unknown function that will be determined by requirement that $\tau_{marg}$ and $r=\mathrm{const}$ solve the equations of motion for $r$.
First of all we see that the new term in (\ref{Bmod}) does not contribute to the equation of motion for $\tau$ since it is equal to 
zero for $\dot{r}$. On the other hand it gives following contribution to the equation of motion for $r$
\begin{eqnarray}\label{eqrmodterm}
\frac{1}{1+\frac{\tau^2}{2}\sqrt{1-2\lambda^{-1}r^2}}
\frac{1}{\sqrt{\bB}}(\partial_rf(r,\tau)r+f(\tau,r))\dot{r}\dot{\tau}
-\frac{d}{d\xi^0}
\left(\frac{1}{1+\frac{\tau^2}{2}\sqrt{1-2\lambda^{-1}r^2}}f(\tau,r)\frac{r\dot{\tau}}{\sqrt{\bB}}\right) \ . \nonumber \\	
\end{eqnarray}
Let us now evaluate this expression when  $\tau=\tau_{marg}$ and when $r=\mathrm{const}$. Then the 
first term in (\ref{eqrmodterm}) is zero while the second one is equal to 
\begin{eqnarray}\label{expmarg}
&&	\frac{d}{d\xi^0}
\left(\frac{r\dot{\tau}}{1+\frac{\tau^2}{2}\sqrt{1-2\lambda^{-1}r^2}}f(\tau,r)\right)=
\nonumber \\
&&=\frac{r\lambda^{-1}\omega^2\left[f(r,\tau)(\tau-\frac{\tau^3}{2}\sqrt{1-2\lambda^{-1}r^2})+\partial_\tau f(\tau^2+\frac{\tau^4}{2}\sqrt{1-2\lambda^{-1}r^2})\right]}
{(1+\frac{\tau^2}{2}\sqrt{1-2\lambda^{-1}r^2})^2} \ . \nonumber \\
\end{eqnarray}
Then in order to obey equation of motion for $r$ we should now demand that (\ref{expmarg}) is equal to
\begin{equation}
\frac{r\tau^4}{2(1+\frac{\tau^2}{2}\sqrt{1-2\lambda^{-1}r^2})^2}
\end{equation}
that implies differential equation for $f$ in the form 
\begin{equation}
f(\tau-\frac{\tau^3}{2}\sqrt{1-2\lambda^{-1}r^2})+\partial_\tau f(\tau^2+\frac{\tau^4}{2}\sqrt{1-2\lambda^{-1}r^2})=\frac{\lambda}{2\omega^2}\tau^4 \ . 
\end{equation}
This equation can be easily solved with result
\begin{eqnarray}
f=\frac{2\lambda}{(1-2\lambda^{-1}r^2)^2}(1+\ln(1+\frac{1}{2}\tau^2\sqrt{1-2\lambda^{-1}r^2})(1+\frac{\tau^2}{2}
\sqrt{1-2\lambda^{-1}r^2})) \ .  \nonumber \\
\end{eqnarray}
This is central result of our analysis since we have shown that it is possible to have
exact form of the action for D0-brane anti-D0-brane system where the marginal tachyon profile at constant relative distance is its exact solution. The crucial point is the presence of the mixed term under square root that is proportional to the function $f(\tau,r)$ given above. Of course, this form of this function is not fixed uniquely since
we can still add to it term which depends on $r$ only. Further, this action is valid for the separation of D0-brane anti-D0-brane below critical distance $r_c=\frac{1}{\sqrt{2\lambda}}$. On the other hand there is nothing wrong with this restriction since the tachyon profile $\tau_{marg}$ is marginal for $r<r_c$ and hence it is natural to define the tachyon effective action for $r<r_c$ only.

\section{Covariant Form of the Action and Dp-Brane Generalization}\label{fourth}
In this section we generalize the form of the proposed action into its covariant form. Recall that 
the action for D0-brane anti-D0-brane system when restricted to the dynamics of relative separation and real tachyon has the form 
\begin{eqnarray}
& &	S=\int d\xi^0 \mL \ ,  \nonumber \\ &&\mL=-\frac{2T_0}{\sqrt{1+\frac{\tau^2}{2}\sqrt{1-2\lambda^{-1}r^2}}}
	\sqrt{1-2F(\tau,r)r \dot{\tau}\dot{r}-\frac{1}{1+\frac{\tau^2}{2}\sqrt{1-2\lambda^{-1}r^2}}\frac{\lambda \dot{\tau}^2}
		{\sqrt{1-2\lambda^{-1}r^2}}} \ , \nonumber \\
\end{eqnarray}
where we  introduced function $F(\tau,r)$ defined as
\begin{equation}
	F(\tau,r)=\frac{f(\tau,r)}{1+\frac{\tau^2}{2}\sqrt{1-2\lambda^{-1}r^2}} \ . 
\end{equation}
Let us introduce coordinates $X^{(i)I},I=1,\dots,9,i=1,2$ that label transverse positions of D0-brane anti-D0-brane respectively. Further, we generalize $r$ to $9-$dimensional vector $r^I$ defined
\begin{equation}
	r^I=X^{(1)I}-X^{(2)I} \ . 
\end{equation}	
It is important to stress that  $\tau$ is real part of the complex tachyon $T$ that is defined as 
\begin{equation}
	T=\tau e^{i\rho} \ ,
\end{equation}
Then it is natural to  replace $\tau^2$ as
\begin{equation}
	\tau^2\rightarrow TT^* \ . 
	\end{equation}
We further generalize expression $2F(\tau)r\dot{\tau}\dot{r}$ into the form
\begin{equation}
F(T)\dot{T}(X^{(1)I}-X^{(2)I})(\dot{X}^{(1)}_I-\dot{X}^{(2)}_I)+
F(T)^*\dot{T}^*(X^{(1)I}-X^{(2)I})(\dot{X}^{(1)}_I-\dot{X}^{(2)}_I) \ , 	
\end{equation}
where $F(T)\equiv\tilde{F}(T^2)\frac{1}{\tau}e^{-i\rho} \ , 
F^*(T)=\tilde{F}(T^2)\frac{1}{\tau}e^{i\rho}
$ 
so that
\begin{eqnarray}
	F(T)\dot{T}+F(T)^*\dot{T}^*=
2\tilde{F}(T^2)\frac{1}{\tau}\dot{\tau} \ .  
\nonumber \\
\end{eqnarray}
finally we replace $\dot{T}$ with covariant derivative
\begin{eqnarray}
&&\dot{T}\Rightarrow D_0T\equiv\partial_t T-i A^{(1)}_0T+iA_0^{(2)}T \ , 
\nonumber \\
&&\dot{T}^*\Rightarrow (D_0T)^*=\partial_t T^*+iA^{(1)}_0 T^*-iA_0^{(2)}T^* \ , 
\nonumber \\ 
\end{eqnarray}
where $A^{(1)}_0$ is the gauge field living on the world-line of D0-brane while
$A_0^{(2)}$ is the gauge field defined on the world-line of anti-D0-brane.

As the final step we introduce kinetic term for $X^{(1)I}$ and $X^{(2)I}$ when, in order to have contact with the  previous forms of the D0-brane anti-D0-brane actions, we introduce these kinetic terms in each own square root function. Then the final form of the action for D0-brane anti-D0-brane system in the form 
\begin{eqnarray}
& &	\mL_{D0+\overline{D0}}=-\frac{T_0}{\sqrt{1+\frac{TT^*}{2}
			\sqrt{1-2\lambda^{-1}(X^{(1)I}-X^{(2)I})(X^{(1)}_I-X^{(2)}_I)}}}
(\sqrt{\bA_{(1)}}+\sqrt{\bA_{(2)}}) \ , \nonumber \\
&&\bA_{(i)}=1-\dot{X}^{(i)I}\dot{X}^{(i)}_{I}-
F(T)D_0T(X^{(1)I}-X^{(2)I})(\dot{X}^{(1)}_{I}-\dot{X}^{(2)}_{I})-
\nonumber \\
&&-F^*(T)D^*_0T(X^{(1)I}-X^{(2)I})(\dot{X}^{(1)}_{I}-\dot{X}^{(2)}_{I})-\nonumber \\
&&-\frac{\lambda}{1+\frac{TT^*}{2}
\sqrt{1-2\lambda^{-1}(X^{(1)I}-X^{(2)I})(X^{(1)}_{I}-X^{(2)}_{I})}}	
D_0 T(D_0T)^* \ .  \nonumber \\	
\end{eqnarray}	
Note that this action is defined in the static gauge that by definition	
\begin{equation}
	X^0_{(1)}=X^{0}_{(2)}\equiv t=\xi^0 \ .
\end{equation}
Leaving the static gauge we replace $\sqrt{\bA_{(i)}}$ with $\sqrt{-\bA_{00}^{(i)}}$ where $\bA_{00}^{(i)}$ is defined as 
\begin{eqnarray}
& &\bA_{00}^{(i)}=\eta_{MN}\partial_0 X^{(i)M}\partial_0 X^{(i)N}+
F(T)D_0T(X^{(1)M}-X^{(2)M})\eta_{MN}(\partial_0X^{(1)N}-\partial_0X^{(2)N})+
\nonumber \\
&&+F^*(T)D^*_0T(X^{(1)M}-X^{(2)M})\eta_{MN}(\partial_0X^{(1)N}-\partial_0X^{(2)N})-\nonumber \\
&&-\frac{\lambda}{1+\frac{TT^*}{2}
	\sqrt{1-2\lambda^{-1}(X^{(1)M}-X^{(2)M})\eta_{MN}(X^{(1)N}-X^{(2)N})}}	
D_0 T(D_0T)^*  \ . 
\nonumber \\
\end{eqnarray}
Note that this form of the action can be straightforwardly extended to the case of $p+1$ dimensional objects which are  Dp-branes    when we replace $\bA_{00}^{(i)}$ with $(p+1)\times (p+1)$ matrix $\bA_{\alpha\beta}^{(i)}$ so that the action has the form
\begin{eqnarray}
&&S=\int d^{p+1}\xi \mL \ , \nonumber \\
&&\mL=-\frac{T_p}{\sqrt{1+\frac{TT^*}{2}
		\sqrt{1-2\lambda^{-1}(X^{(1)M}-X^{(2)M})\eta_{MN}(X^{(1)N}-X^{(2)N})}}}\times 
	\nonumber \\
	&&\times 
\left(\sqrt{-\det\bA_{\alpha
\beta}^{(1)}}+\sqrt{-\det\bA_{\alpha\beta}^{(i)}}\right) \ , \nonumber \\
&&\bA_{\alpha\beta}^{(i)}=\eta_{MN}\partial_\alpha X^{(i)M}\partial_\beta X^{(i)N}+
F(T)D_\alpha T(X^{(1)M}-X^{(2)M})\eta_{MN}(\partial_\beta X^{(1)N}-\partial_\beta X^{(2)N})+
\nonumber \\
&&+F^*(T)(D_\alpha T)^*(X^{(1)M}-X^{(2)M})\eta_{MN}(\partial_\beta X^{(1)N}-\partial_\beta X^{(2)N})+\lambda F_{\alpha\beta}^{(i)}-\nonumber \\
&&-\frac{\lambda}{1+\frac{TT^*}{2}
	\sqrt{1-2(X^{(1)M}-X^{(2)M})\eta_{MN}(X^{(1)N}-X^{(2)N})}}	
\frac{1}{2}(D_\alpha T(D_\beta T)^*+D_\beta T (D_\alpha T)^*) \ . 
\nonumber \\
\end{eqnarray}
This is the final form of the action for Dp-brane anti-Dp-brane system. Note that the matrix $\bA_{\alpha\beta}$ contain  terms with  derivatives of $X$ and $T$ that have similar structure as in the action derived in 
\cite{Garousi:2004rd}. Then one can ask the question whether there exists field redefinition that maps these two tachyon effective actions each other. This problem is currently under study.

{\bf Acknowledgement:}
\\
This work 
is supported by the grant “Integrable Deformations”
(GA20-04800S) from the Czech Science Foundation
(GACR). 



\end{document}